\documentclass[11pt]{article}
\usepackage[dvips]{graphicx}
\usepackage{amsmath,amssymb}
\usepackage{amscd}
\usepackage[all]{xy}

\newtheorem{Def}{Definition}
\newtheorem{Thm}[Def]{Theorem}
\newtheorem{Lem}[Def]{Lemma}
\newtheorem{Pro}[Def]{Proposition}
\newtheorem{Cor}[Def]{Corollary}

\newcommand{\To}{\Rightarrow}

\newcommand{\Tot}{\Leftrightarrow}

\begin{document}
\title{Permutation Complexity via Duality between Values and Orderings}

\author{Taichi Haruna\footnote{Corresponding author}$\ ^{\rm ,1,2}$,  Kohei Nakajima$\ ^{\rm 3}$ \\
\footnotesize{$\ ^{\rm 1}$ Department of Earth \& Planetary Sciences, Graduate School of Science, } \\
\footnotesize{Kobe University, 1-1, Rokkodaicho, Nada, Kobe, 657-8501, JAPAN} \\
\footnotesize{$\ ^{\rm 2}$ PRESTO, Japan Science and Technology Agency (JST),} \\
\footnotesize{4-1-8 Honcho Kawaguchi, Saitama 332-0012, Japan} \\
\footnotesize{$\ ^{\rm 3}$ Artificial Intelligence Laboratory, Department of Informatics, } \\
\footnotesize{University of Zurich, Andreasstrasse 15, 8050 Zurich, Switzerland} \\
\footnotesize{E-mail: tharuna@penguin.kobe-u.ac.jp (T. Haruna)} \\
\footnotesize{Tel \& Fax: +81-78-803-5739} \\
}

\date{}
\maketitle

\begin{abstract}
We study the permutation complexity of finite-state stationary stochastic processes based on a 
duality between values and orderings between values. First, we establish a duality 
between the set of all words of a fixed length and the set of all permutations of the same length. 
Second, on this basis, we give an elementary alternative proof of the equality between the permutation entropy rate 
and the entropy rate for a finite-state stationary stochastic processes first proved in 
[Amig\'o, J.M., Kennel, M. B., Kocarev, L., 2005. {\it Physica D} 210, 77-95]. 
Third, we show that further information on the relationship between the structure of values and 
the structure of orderings for finite-state stationary stochastic processes beyond the entropy rate 
can be obtained from the established duality. 
In particular, we prove that the permutation excess entropy is equal to the excess entropy, 
which is a measure of global correlation present in a stationary stochastic process, for 
finite-state stationary ergodic Markov processes. 
\end{abstract}


\section{Introduction}
One of the most intriguing recent findings in the science of complexity is 
that much of the information contained in stationary time series can be captured 
by orderings between values \cite{Amigo2010}. Bandt and Pompe \cite{Bandt2002a} first introduced the notion of 
permutation entropy which quantifies the average uncertainty of orderings 
between values per time unit, in contrast to the entropy rate for stationary 
stochastic processes or the Kolmogorov-Sinai entropy for dynamical systems, 
both of which quantify the average uncertainty of values per time unit. 
Bandt et al. \cite{Bandt2002b} proved that the permutation entropy is equal to the Kolmogorov-Sinai 
entropy for piecewise monotone maps on one-dimensional intervals. Amig\'o et al. 
\cite{Amigo2005} showed that the permutation entropy rate is equal to the entropy rate for any 
finite-state stationary stochastic process
\footnote{
Amig\'o et al. stated that the equality holds for finite-state stationary ergodic processes in 
Theorem 2 and an inequality holds for the non-ergodic case in Theorem 6 in \cite{Amigo2005}. 
However, one can see that they actually proved the equality for any 
finite-state stationary stochastic process if he or she examine their proof carefully. 
This point is corrected in Amig\'o's recent book \cite{Amigo2010}. 
}. 
They also generalized the results 
of \cite{Bandt2002b} to ergodic maps on intervals of arbitrary dimensions by considering 
the limits of finite-state stationary stochastic processes. Keller and Sinn \cite{Keller2010} took a 
different approach from that of \cite{Amigo2005} to generalize the results of \cite{Bandt2002b}. 
The topological permutation entropy was also studied by Bandt et al. \cite{Bandt2002b}, 
Misiurewicz \cite{Misiurewicz2003} and Amig\'o and Kennel \cite{Amigo2007}. 

In this paper, we study the permutation complexity of finite-state 
stationary stochastic processes based on a duality between values and orderings between values. 
Orderings between values induce a coarse-graining of the set of all words of a fixed length. Namely, 
two words are mapped to the same ordering (permutation) if order-relationships between values in 
both words are the same. 
In the case of shift maps on the unit interval, Elizalde \cite{Elizalde2009} performed enumerations 
associated with such a coarse-graining. In our case, the enumeration is similar, but 
much simpler than that of \cite{Elizalde2009}. However, we emphasize a dual 
structure existing between the set of all words of a fixed length and the set of all permutations 
of the same length. Indeed, we show that there is a kind of minimal realization map 
from the latter to the former. We can make the pair of the coarse-graining map and the minimal realization map 
form a Galois connection \cite{Davey2002}, which is a categorical adjunction \cite{MacLane1998} 
between partially ordered sets, by introducing suitable partial orders on the sets at both sides. 
We present an elementary alternative proof for the equality between the permutation entropy rate and 
the entropy rate based on the duality between values and orderings. 

We can study the further relationship between the structure of values and the structure of orderings 
for finite-state stationary stochastic processes beyond the entropy rate equality if we make use of the 
duality between values and orderings in more depth. Here, we consider the excess entropy which is a measure of global 
correlation present in finite-state stationary stochastic processes. The excess entropy has an old 
history in complex systems study \cite{Crutchfield1983, Grassberger1986,Shaw1984}. 
However, it is still of recent research interest. For example, Feldman et al. \cite{Feldman2008} 
proposed the entropy-complexity diagrams based on the entropy rate 
and the excess entropy to analyze various types of natural information processing. We define the permutation 
excess entropy and show that the permutation excess entropy is equal to the excess entropy for 
finite-state stationary ergodic Markov processes. We also present a simple non-ergodic counter-example 
with a strict inequality. 

Let us give a rough sketch of our proof strategy for the main results. 
Let $\phi$ be the coarse-graining map sending each word of length $L(\geq 1)$ from a finite alphabet 
to its associated permutation of length $L$. Given a finite-state stationary stochastic process, 
only permutations $\pi$ such that the size of $\phi^{-1}(\pi)$ is greater than 1 may contribute to the difference between 
the entropy rate and the permutation entropy rate of the process. If we denote the probability that those permutations 
occur by $q_L$, then we can show that the difference ($\geq 0$) before the normalization (division by $L$) and 
taking the limit of $L \to \infty$ is bounded from above by the probability $q_L$ multiplied by a function of 
$L$ whose growth rate is $\log L$ by using the fact that the size of $\phi^{-1}(\pi)$ is given by a binomial 
coefficient depending on $L$ for any permutation $\pi$ of length $L$ (Lemma \ref{lem3}). 
The equality between the entropy rate and the permutation entropy rate is immediate from this bound (Theorem \ref{thm3}). 
Furthermore, if the process is ergodic Markov, 
then we can show that $q_L$ diminishes exponentially fast as $L \to \infty$ by using a characterization of 
words $s_1^L$ such that $\phi^{-1}(\pi)=\{s_1^L\}$ for some $\pi$ and the irreducibility of the 
associated transition matrix. This leads to the equality between the excess entropy and 
the permutation excess entropy (Theorem \ref{thm2}). We note that those words $s_1^L$ such that 
$\phi^{-1}(\pi)=\{s_1^L\}$ for some $\pi$ can be seen as a special type of ``stable objects'' under 
the duality between the coarse-graining map $\phi$ and the minimal realization map (Theorem \ref{thm1} (iii)). 

This paper is organized as follows. 
In Section 2, we establish the duality between values and orderings. 
In Section 3, we give a proof of the equality between the permutation entropy rate and the entropy rate 
for finite-state stationary stochastic processes based on the duality. 
In Section 4, we prove the equality between the permutation excess entropy and the excess entropy 
for finite-state stationary ergodic Markov processes and give a non-ergodic counter-example with a 
strict inequality. 

\section{Duality between Values and Orderings}
In this section we establish the duality between values and orderings. 

\subsection{Permutations and Rank Sequences}
Let $A$ be an alphabet. We consider the case that the cardinality $|A|$ of $A$ is finite or countably infinite. 
If $|A|=n \ (n=1,2,\cdots)$, then we write $A=A_n=\{1,2,\cdots, n\}$. 
If $n=\infty$, then $A=A_\infty$ is identified with the set of all natural numbers 
$\mathbb{N}=\{1,2,3,\cdots \}$. We consider $A_n \ (n=1,2,\cdots,\infty)$ is not just a set, but a totally ordered 
set ordered by the `less-than-or-equal-to' relationship $\leq$ between natural numbers. 
In the following discussion, if we write just $A$, then $A$ can be either $A_n$ or $A_\infty=\mathbb{N}$. 

Let $A^L=\underbrace{A \times \cdots \times A}_{L}$ for $L \geq 1$. Each element $w \in A^L$ is called 
a {\it word} of length $L$. If $w=(s_1,\cdots,s_L) \in A^L$, then we write $w=s_1 \cdots s_L=s_1^L$. 

Let $\mathcal{S}_L$ be the set of all {\it permutations} of length $L$, namely, 
$\mathcal{S}_L$ is the set of all bijections on the set $\{1,2,\cdots, L\}$. 
For $s_1^L \in A^L$ and $\pi \in \mathcal{S}_L$, 
we say that $s_1^L$ is of type $\pi$ if we have 
$s_{\pi(i)} \leq s_{\pi(i+1)}$ and $\pi(i) < \pi(i+1)$ when $s_{\pi(i)} = s_{\pi(i+1)}$ 
for $i =1,2,\cdots,L-1$. 
For example, $\pi(1)\pi(2)\pi(3)\pi(4)\pi(5)=24315$ for $s_1^5=31213$ because 
$s_2 s_4 s_3 s_1 s_5=11233$. 

Each word $s_1^L \in A^L$ has a unique permutation type $\pi \in \mathcal{S}_L$. 
Hence, the correspondence $s_1^L \mapsto \pi$ defines a many-to-one (in general) map $\phi: A^L \to \mathcal{S}_L$, which 
coarse-grains the set $A^L$ of words of length $L$ by their permutation types. 

We make use of the notion of {\it rank sequence} introduced in \cite{Amigo2005}. 
In some situations, discussions might become facilitated if we use rank sequences instead of permutations. 
However, as far as the authors are aware, their compatibility with the map $\phi$ sending words to associated permutations 
has not been presented explicitly so far. Hence, it may not be worthless to study them here. 

A word $r_1^L \in \mathbb{N}^L$ is called a {\it rank sequence} of length $L$ if it satisfies 
$1 \leq r_i \leq i$ for $i=1,\cdots,L$. We denote the set of all rank sequences of length $L$ 
by $\mathcal{R}_L$. Note that there exists a bijection between $\mathcal{S}_L$ and $\mathcal{R}_L$ 
because $|\mathcal{S}_L|=L!=|\mathcal{R}_L|$. 

Each word $s_1^L \in A^L$ gives rise to a rank sequence $r_1^L \in \mathcal{R}_L$ in the following way:
\begin{eqnarray*}
r_i=\sum_{j=1}^i \delta(s_j \leq s_i), \ i=1,\cdots,L, 
\end{eqnarray*}
where $\delta(X)=1$ if the proposition $X$ is true, otherwise $\delta(X)=0$. Namely, 
$r_i$ is the number of indices $j \ (1 \leq j \leq i)$ such that $s_j \leq s_i$. 
This correspondence $s_1^L \mapsto r_1^L$ defines a map $\varphi:A^L \to \mathcal{R}_L$. 

In the following discussion, we will show that there is a bijection 
$\iota: \mathcal{R}_L \to \mathcal{S}_L$ such that $\iota \circ \varphi=\phi$, 
namely, the following diagram commutes: 
\begin{eqnarray*}
\xymatrix{
A^L \ar[r]^{\phi} \ar[dr]_{\varphi} & \mathcal{S}_L \\
& \mathcal{R}_L. \ar[u]_{\iota}
}
\end{eqnarray*}
Given a rank sequence $r_1^L \in \mathcal{R}_L$, we define a permutation $\iota(r_1^L)=\pi \in \mathcal{S}_L$ 
inductively as follows: 
first, we define $\pi(1)=\max\{i | r_i=1, \ 1 \leq i \leq L\}$. $\pi(1)$ is well-defined because 
we have $r_1=1$. Second, we define 
\begin{eqnarray*}
\pi(2)=\max \{ i | r_i^{(1)}=\min_{j \neq \pi(1)} \{r_j^{(1)}\} \}
\end{eqnarray*}
where $r_1^{(1)} \cdots r_L^{(1)}$ is a rank sequence defined by
\begin{eqnarray*}
r_i^{(1)}=
\begin{cases}
r_i-1 & \text{if} \ i > \pi(1) \\
r_i & \text{otherwise.}
\end{cases}
\end{eqnarray*}
In general, we define 
\begin{eqnarray*}
\pi(k)=\max \{ i | r_i^{(k-1)}=\min \{r_j^{(k-1)} | j \neq \pi(1), \cdots,\pi(k-1) \} \}
\end{eqnarray*}
for $k=2,\cdots,L$, where $r_1^{(k-1)} \cdots r_L^{(k-1)}$ is a rank sequence defined by 
\begin{eqnarray*}
r_i^{(k-1)}=
\begin{cases}
r_i^{(k-2)}-1 & \text{if} \ i > \pi(k-1) \ \text{and} \ i \neq \pi(1),\cdots,\pi(k-2) \\
r_i^{(k-2)} & \text{otherwise,}
\end{cases}
\end{eqnarray*}
and $r_i^{(0)}=r_i$. By construction, this procedure defines a unique permutation 
$\iota(r_1^L)=\pi \in \mathcal{S}_L$. 

For example, consider $r_1^5=11342 \in \mathcal{R}_5$. $\pi=\iota(11342) \in \mathcal{S}_5$ is obtained by the 
following calculations: 
\begin{eqnarray*}
\pi(1)&=&\max \{ i | r_i=1 \}=2, \ r_1^{(1) 5}=11231, \\
\pi(2)&=&\max \{ i | r_i^{(1)}=\min_{j \neq 2} \{r_j^{(1)}\} \}=5, \ r_1^{(2) 5}=11231, \\
\pi(3)&=&\max \{ i | r_i^{(2)}=\min_{j \neq 2,5} \{r_j^{(2)}\} \}=1, \ r_1^{(3) 5}=11121, \\
\pi(4)&=&\max \{ i | r_i^{(3)}=\min_{j \neq 1,2,5} \{r_j^{(3)}\} \}=3, \ r_1^{(4) 5}=11111, \\
\pi(5)&=&\max \{ i | r_i^{(4)}=\min_{j \neq 1,2,3,5} \{r_j^{(4)}\} \}=4. 
\end{eqnarray*}

\begin{Lem}
\begin{eqnarray*}
r_{\pi(k)}^{(k-1)}=1
\end{eqnarray*}
for $k=1,2,\cdots,L$. 
\label{lem1}
\end{Lem}
{\it Proof.}
It is sufficient to show that $r_j^{(k-1)}=1$ for some $j \neq \pi(1),\cdots,\pi(k-2)$. 
Consider the minimum index $j$ such that $j \not \in \{ \pi(1),\cdots,\pi(k-2) \}$. 
Then, we have $r_j^{(k-1)}=r_j-(j-1)$ because $\{1,\cdots,j-1\} \subseteq \{\pi(1), \cdots,\pi(k-2)\}$. 
However, $1 \leq r_j \leq j$ and $r_j^{(k-1)} \geq 1$ by construction. 
Hence, $r_j=j$ and we obtain $r_j^{(k-1)}=1$.

\hfill $\Box$ \\

\begin{Pro}
The map $\iota : \mathcal{R}_L \to \mathcal{S}_L$ is a bijection. 
\label{pro1}
\end{Pro}
{\it Proof.}
It is sufficient to show that $\iota$ is injective because $|\mathcal{R}_L|=|\mathcal{S}_L|=L!<\infty$. 
Assume that $\iota(r_1^L)=\iota(\tilde{r}_1^L)=\pi$ for $r_1^L, \tilde{r}_1^L \in \mathcal{R}_L, \ \pi \in \mathcal{S}_L$. 
We have $r_{i}^{(L-1)}=\tilde{r}_{i}^{(L-1)}=1$ for $i=1,\cdots,L$ by Lemma \ref{lem1} 
because $r_{\pi(k)}^{(k-1)}=r_{\pi(k)}^{(L-1)}$ for $k=1,\cdots,L$. 
We can reconstruct both $r_1^L$ and $\tilde{r}_1^L$ from 
$\overline{r}_1^{(L-1)L}:=r_{1}^{(L-1)L}=\tilde{r}_{1}^{(L-1)L}=\underbrace{11\cdots1}_{L}$ by the following procedure: 
first, we add $1$ to the $\pi(L)$-th $1$ in $\overline{r}_1^{(L-1)L}$ 
if $\pi(L)>\pi(L-1)$, and do nothing otherwise. The obtained sequence $\overline{r}_1^{(L-2)L}$ is identical to 
both $r_1^{(L-2)L}$ and $\tilde{r}_1^{(L-2)L}$ because $\iota(r_1^L)=\iota(\tilde{r}_1^L)=\pi$. 
Second, we add $1$ to $\overline{r}_{\pi(L)}^{(L-2)}$ if $\pi(L)>\pi(L-2)$, and do nothing otherwise, 
and add $1$ to $\overline{r}_{\pi(L-1)}^{(L-2)}$ if $\pi(L-1)>\pi(L-2)$, and do nothing otherwise. 
If we call the obtained sequence 
$\overline{r}_{1}^{(L-3)L}$, then we have $\overline{r}_{1}^{(L-3)L}=r_1^{(L-3)L}=\tilde{r}_1^{(L-3)L}$. In general, 
if we define 
\begin{eqnarray*}
\overline{r}_i^{(L-k)}=
\begin{cases}
\overline{r}_i^{(L-(k-1))}+1 & \text{if} \ i \in \{ \pi(L-(k-1)), \cdots, \pi(L) \} \ \text{and} \ i > \pi(L-k) \\
\overline{r}_i^{(L-(k-1))} & \text{otherwise} 
\end{cases}
\end{eqnarray*}
for $k=2,\cdots,L$, then we have $\overline{r}_{1}^{(L-k)L}=r_1^{(L-k)L}=\tilde{r}_1^{(L-k)L}$. In particular, we obtain 
$\overline{r}_{1}^{(0)L}=r_1^L=\tilde{r}_1^L$ for $k=L$. 

\hfill $\Box$ \\

\begin{Pro}
$\iota \circ \varphi=\phi$. 
\label{pro2}
\end{Pro}
{\it Proof.}
We have to show that $\iota(\varphi(s_1^L))=\phi(s_1^L)$ for any word $s_1^L \in A^L$. 
Put $\pi=\phi(s_1^L)$, $\tilde{\pi}=\iota(\varphi(s_1^L))$ and $r_1^L=\varphi(s_1^L)$. 
We shall show that $\pi(k)=\tilde{\pi}(k)$ for $k=1,\cdots,L$ inductively. 
First, we show that $\pi(1)=\tilde{\pi}(1)$. 
By the definition of $\phi$ and $\iota$, $\pi(1)$ is the index $i$ of the minimum-leftmost $s_i$ and 
$\tilde{\pi}(1)$ is the maximum index $i$ such that $r_i=1$. 
We have 
\begin{eqnarray*}
r_i=1 \Tot s_j > s_i \ \text{for} \ j=1,\cdots,i-1
\end{eqnarray*}
by the definition of rank sequences. Hence, $s_j>s_{\tilde{\pi}(1)}$ for $j=1,\cdots,\tilde{\pi}(1)-1$.
On the other hand, we have $s_{\tilde{\pi}(1)} \leq s_j$ for $j=\tilde{\pi}(1), \cdots, L$. 
Indeed, if there exists $j > \tilde{\pi}(1)$ such that $s_{\tilde{\pi}(1)}> s_j$, then 
$r_j > 1$ must hold because $\tilde{\pi}(1)$ is the maximum index $i$ such that $r_i=1$. 
Hence, there exists $j_1 < j$ such that $s_{j_1} \leq s_j$. If $j_1 \leq \tilde{\pi}(1)$, then 
this contradicts $s_k>s_{\tilde{\pi}(1)}$ for $k=1,\cdots,\tilde{\pi}(1)-1$. So, 
we have $\tilde{\pi}(1)<j_1<j$. Since $s_{\tilde{\pi}(1)}> s_j \geq s_{j_1}$, the same argument 
can be applied to $j_1$ instead of $j$. Thus, we obtain a strictly decreasing infinite sequence of indices 
$j_1 j_2 \cdots$ such that $\tilde{\pi}(1) < \cdots < j_2 < j_1 < j$. However, this is impossible because 
the number of indices between $\tilde{\pi}(1)$ and $j$ is finite. 
Therefore, $s_{\tilde{\pi}(1)}$ is the minimum-leftmost value in $s_1^L$, which implies that 
$\tilde{\pi}(1)=\pi(1)$. 

Now, suppose that $\tilde{\pi}(1)=\pi(1),\cdots,\tilde{\pi}(k)=\pi(k)$, where $1 \leq k \leq L-1$. 
We would like to show that $\tilde{\pi}(k+1)=\pi(k+1)$. 
By the definition of $\phi$ and $\iota$, $\pi(k+1)$ is the index $i$ of the minimum-leftmost $s_i$ except 
for $s_{\pi(1)},\cdots,s_{\pi(k)}$ and 
$\tilde{\pi}(k+1)$ is the maximum index $i$ such that $r_i^{(k)}=1$ except for $\tilde{\pi}(1),\cdots,\tilde{\pi}(k)$, 
where we have $\tilde{\pi}(1)=\pi(1),\cdots,\tilde{\pi}(k)=\pi(k)$ by the assumption of the mathematical induction. 
For an appropriate permutation $(i_1,\cdots,i_k)$ of $(1,\cdots,k)$, we have 
\begin{eqnarray*}
\pi(i_1) < \cdots < \pi(i_m)< \tilde{\pi}(k+1) < \pi(i_{m+1}) < \cdots < \pi(i_k).
\end{eqnarray*}
It must hold that $r_{\tilde{\pi}(k+1)}=m+1$ because $r_{\tilde{\pi}(k+1)}^{(k)}=1$. 
The number of indices $j$ for $j=1,\cdots,\tilde{\pi}(k+1)-1$ such that $s_j \leq s_{\tilde{\pi}(k+1)}$ is $m$ 
by the definition of $r_{\tilde{\pi}(k+1)}$. On the other hand, we have 
$s_{\pi(i_1)}, \cdots, s_{\pi(i_m)} \leq s_{\tilde{\pi}(k+1)}$ by the definition of $\pi$. Hence, the equality 
\begin{eqnarray*}
\{ j | s_j \leq s_{\tilde{\pi}(k+1)} \ \text{and} \ 1 \leq j < \tilde{\pi}(k+1) \}
=\{ \pi(i_1), \cdots, \pi(i_m) \}
\end{eqnarray*}
holds. Thus, if $j \neq \pi(i_1),\cdots,\pi(i_m)$ and $1 \leq j < \tilde{\pi}(k+1)$, then 
we have $s_j > s_{\tilde{\pi}(k+1)}$. This implies that $\tilde{\pi}(k+1) \leq \pi(k+1)$ because 
if $\pi(k+1) < \tilde{\pi}(k+1)$, then $s_{\pi(k+1)} > s_{\tilde{\pi}(k+1)}$, which contradicts 
the assumption that $s_{\pi(k+1)}$ takes the minimum value except for $s_{\pi(1)},\cdots,s_{\pi(k)}$. 
For the other inequality, assume that $\pi(i_{m'}) < \pi(k+1) <\pi(i_{m'+1})$. 
We have $s_j > s_{\pi(k+1)}$ for $j \neq \pi(i_1),\cdots,\pi(i_{m'})$ because $s_{\pi(k+1)}$ takes the 
minimum-leftmost value except for $s_{\pi(1)},\cdots,s_{\pi(k)}$. On the other hand, 
it follows that $s_{\pi(i_1)},\cdots, s_{\pi(i_{m'})} \leq s_{\pi(k+1)}$ by the definition of $\pi$. 
Hence, we have $r_{\pi(k+1)}=\sum_{j=1}^{\pi(k+1)} \delta(s_j \leq s_{\pi(k+1)})=m'+1$, which implies that 
$r_{\pi(k+1)}^{(k)}=1$. Thus, we obtain $\pi(k+1) \leq \tilde{\pi}(k+1)$ because 
$\tilde{\pi}(k+1)$ is the maximum index $i$ such that $r_i^{(k)}=1$ except for $\pi(1),\cdots,\pi(k)$.

\hfill $\Box$ \\

\begin{Cor}
For $s_1^L, t_1^L \in A^L$, the following statements are equivalent: 
\begin{itemize}
\item[(i)]
$\phi(s_1^L)=\phi(t_1^L)$. 
\item[(ii)]
For all $1 \leq j \leq k \leq L$, $s_k \leq s_j \Tot t_k \leq t_j$. 
\end{itemize}
\label{cor1}
\end{Cor}
{\it Proof.}
Assume that $\phi(s_1^L)=\phi(t_1^L)=\pi \in \mathcal{S}_L$. Then, we have 
\begin{eqnarray*}
s_{\pi(1)} \leq s_{\pi(2)} \leq \cdots \leq s_{\pi(L)} \ \text{and} \\
t_{\pi(1)} \leq t_{\pi(2)} \leq \cdots \leq t_{\pi(L)}. 
\end{eqnarray*}
Hence, (ii) holds. 
For the reverse direction, assume that (ii) holds. Then, we have 
$\sum_{k=1}^j \delta(s_k \leq s_j)=\sum_{k=1}^j \delta(t_k \leq t_j)$ for 
any $1 \leq j \leq L$, which implies $\varphi(s_1^L)=\varphi(t_1^L)$. 
Hence, we have $\phi(s_1^L)=\iota \circ \varphi(s_1^L)=\iota \circ \varphi(t_1^L)=\phi(t_1^L)$. 

\hfill $\Box$ \\

\subsection{The Coarse-Graining Map $\phi$}
Now, we are ready to study properties of the coarse-graining map $\phi:A^L \to {\mathcal S}_L$ in detail. 

\begin{Lem}
Let $\pi \in \mathcal{S}_L$. 
Assume that there is no $s_1^L \in A_{i-1}^L$ such that $\phi(s_1^L)=\pi$, but 
there exists $s_1^L \in A_i^L$ such that $\phi(s_1^L)=\pi$ for some $i \geq 1$ 
(when $i=1$ we define $A_{i-1}=A_0=\emptyset$). 
\begin{itemize}
\item[(i)]
There exists a unique $s_1^L \in A_i^L$ such that $\phi(s_1^L)=\pi$. 
Moreover, if $\phi(t_1^L)=\pi$ for $t_1^L \in A_n^L$ and $n \geq i$, then 
there exist $c_1,\cdots,c_L$ such that 
$s_k+c_k=t_k$ for $k=1,\cdots,L$ and $0 \leq c_{\pi(1)} \leq \cdots \leq c_{\pi(L)} \leq n-i$. 
\item[(ii)]
$|\phi^{-1}(\pi)|=\binom{L+n-i}{n-i}$, where $n \geq i$ and the domain of $\phi$ is set to $A_n^L$. 
\end{itemize}
\label{lem2}
\end{Lem}
{\it Proof.}
(i): First, we prove the uniqueness. 
If $i=1$, then we have nothing to do. So, we assume that $i \geq 2$. 
Suppose that $\phi(s_1^L)=\phi(t_1^L)=\pi$ and $s_1^L, t_1^L \in A_i^L$. 
If $s_1^L \neq t_1^L$, then there exists $j$ such that 
$s_{\pi(j)} \neq t_{\pi(j)}$. We can assume that $s_{\pi(j)} < t_{\pi(j)}$ 
without loss of generality. Let us define a word $u_1^L$ by 
\begin{eqnarray*}
u_{\pi(k)}=
\begin{cases}
s_{\pi(k)} & k=1,\cdots, j-1, \\
t_{\pi(k)}-1 & k=j,\cdots, L.
\end{cases}
\end{eqnarray*}
We claim that $\phi(u_1^L)=\pi$. 
Indeed, it is clear that we have $u_{\pi(k-1)} \leq u_{\pi(k)}$ and $\pi(k-1) < \pi(k)$ 
when $u_{\pi(k-1)} = u_{\pi(k)}$, for $k \neq j$. 
When $k=j$, we have $s_{\pi(j-1)} \leq s_{\pi(j)} \leq t_{\pi(j)}-1$ by the assumption. 
Suppose that $s_{\pi(j-1)}=t_{\pi(j)}-1$. It follows that $s_{\pi(j-1)}=s_{\pi(j)}$, which implies 
that $\pi(j-1) < \pi(j)$. Thus, we have $\phi(u_1^L)=\pi$. 
However, this contradicts the assumption that there is no $s_1^L \in A_{i-1}^L$ 
such that $\phi(s_1^L)=\pi$ because $u_1^L \in A_{i-1}^L$. 

Next, suppose that $\phi(t_1^L)=\pi$ for $t_1^L \in A_n^L, \ n \geq i$. 
Let us show that $s_{\pi(k)} \leq t_{\pi(k)}$ for $k=1,\cdots,L$. 
If $i=1$, then we have nothing to do because $s_{\pi(k)}=1$ for all $k$. 
So, we assume that $i \geq 2$. 
If there exists $j$ such that $s_{\pi(j)} > t_{\pi(j)}$, then 
a word $u_1^L$ defined by 
\begin{eqnarray*}
u_{\pi(k)}=
\begin{cases}
t_{\pi(k)} & k=1,\cdots, j-1, \\
s_{\pi(k)}-1 & k=j,\cdots, L.
\end{cases}
\end{eqnarray*}
satisfies $\phi(u_1^L)=\pi$ and $u_1^L \in A_{i-1}^L$ by the same reason in the proof 
of the uniqueness, which violates the assumption that there is no $s_1^L \in A_{i-1}^L$ 
such that $\phi(s_1^L)=\pi$. 
Hence, if we define $c_k=t_k-s_k$ for $k=1,\cdots,L$, then $c_k \geq 0$ and 
$c_{\pi(L)} \leq n-i$ because $t_{\pi(L)} \leq n$ and $s_{\pi(L)}=i$. 
The remaining task for us is to show that $c_{\pi(k)} \leq c_{\pi(k+1)}$ for 
$k=1,\cdots,L-1$. 
If $i=1$, then $c_{\pi(k)}=t_{\pi(k)}-1 \leq t_{\pi(k+1)}-1=c_{\pi(k+1)}$ for $k=1,\cdots,L-1$. 
Suppose that $i \geq 2$ and $c_{\pi(j+1)}<c_{\pi(j)}$ for some $j$. 
Then, we have
\begin{eqnarray*}
c_{\pi(j+1)} < c_{\pi(j)} &\Tot& t_{\pi(j+1)}-s_{\pi(j+1)} < t_{\pi(j)}-s_{\pi(j)} \\
&\Tot& s_{\pi(j)}+\left( t_{\pi(j+1)}-t_{\pi(j)} \right) < s_{\pi(j+1)}. 
\end{eqnarray*}
This implies that 
\begin{eqnarray}
s_{\pi(j)} \leq s_{\pi(j)}+\left( t_{\pi(j+1)}-t_{\pi(j)} \right) \leq s_{\pi(j+1)}-1 
\label{eq1}
\end{eqnarray}
because $t_{\pi(j+1)} \geq t_{\pi(j)}$. Let us introduce a word $u_1^L$ by 
\begin{eqnarray*}
u_{\pi(k)}=
\begin{cases}
s_{\pi(k)} & k=1,\cdots, j, \\
s_{\pi(k)}-1 & k=j+1,\cdots, L.
\end{cases}
\end{eqnarray*}
We claim that $\phi(u_1^L)=\pi$ and $u_1^L \in A_{i-1}^L$, which contradicts 
the assumption that there is no $s_1^L \in A_{i-1}^L$ such that $\phi(s_1^L)=\pi$. 
We only need to show that $\pi(j) < \pi(j+1)$ when $u_{\pi(j)}=u_{\pi(j+1)}$. 
However, by (\ref{eq1}), if $s_{\pi(j)}=s_{\pi(j+1)}-1$, then we have 
$t_{\pi(j)}=t_{\pi(j+1)}$, which implies that $\pi(j) < \pi(j+1)$. 

(ii): The number of sequences $c_1 \cdots c_L$ satisfying 
$0 \leq c_{\pi(1)} \leq \cdots \leq c_{\pi(L)} \leq n-i$ 
is given by a binomial coefficient $\binom{L+n-i}{n-i}$. 
Hence, we have $|\phi^{-1}(\pi)| \leq \binom{L+n-i}{n-i}$ by (i). 
On the other hand, given a sequence $c_1 \cdots c_L$ such that 
$0 \leq c_{\pi(1)} \leq c_{\pi(2)} \leq \cdots \leq c_{\pi(L)} \leq n-i$, 
$t_1^L \in A_n^L$ defined by $t_k=s_k+c_k$ for $k=1,\cdots,L$ clearly satisfies $\phi(t_1^L)=\pi$. 
Hence, we have $|\phi^{-1}(\pi)| \geq \binom{L+n-i}{n-i}$. 

\hfill $\Box$ \\

If there is no word $s_1^L \in A_{i-1}^L$ such that $\phi(s_1^L)=\pi$, but 
there exists a (unique) word $s_1^L \in A_i^L$ such that $\phi(s_1^L)=\pi$ for $i \geq 1$, then 
we say that {\it $\pi$ appears for the first time at $i$}. 
We denote the number of permutations $\pi \in \mathcal{S}_L$ that appear for the first time at $i$ 
by $\nu(i,L)$. By Lemma \ref{lem2}, we have $\nu(1,L)=1$ and 
\begin{eqnarray}
\nu(n,L)=n^L- \sum_{i=1}^{n-1} \binom{L+n-i}{n-i} \nu(i,L) 
\label{eq2}
\end{eqnarray}
for $n \geq 2$. 

The following Proposition \ref{pro3} and the subsequent paragraph in this subsection 
are only for the record. They will not be used in later sections. So, readers who are interested in only 
the main results of this paper can skip them. 

\begin{Pro}
A closed-form expression for $\nu(n,L)$ is given by the following formula: 
\begin{eqnarray}
\nu(n,L)=\sum_{i=0}^{n-1} (-1)^i \binom{L+1}{i}(n-i)^L. 
\label{eq3}
\end{eqnarray}
\label{pro3}
\end{Pro}
{\it Proof.}
We prove the formula by mathematical induction on $n$. 
if $n=1$, then we have $\nu(1,L)=1$. 
Assume that the formula holds for natural numbers $1,2,\cdots,n$. Then, we have 
\begin{eqnarray*}
\nu(n+1,L) &=& (n+1)^L-\sum_{i=1}^{n} \binom{L+n+1-i}{n+1-i} \nu(i,L) \\
&=& (n+1)^L-\sum_{i=1}^{n} \binom{L+n+1-i}{n+1-i} \sum_{k=0}^{i-1} (-1)^k \binom{L+1}{k}(i-k)^L \\
&=& (n+1)^L+\sum_{j=1}^{n} \left( \sum_{\begin{subarray}{c} i-k=j, \\ 1 \leq i \leq n \end{subarray}}(-1)^{k+1} \binom{L+n+1-i}{n+1-i}\binom{L+1}{k} \right) j^L. 
\end{eqnarray*}
It is enough to show that 
\begin{eqnarray*}
\sum_{\begin{subarray}{c} i-k=j, \\ 1 \leq i \leq n \end{subarray}}(-1)^{k+1} \binom{L+n+1-i}{n+1-i}\binom{L+1}{k} =(-1)^{n+1-j} \binom{L+1}{n+1-j} 
\end{eqnarray*}
for $j=1,\cdots,n$. 
If we put $l=n-j$, then this is equivalent to showing that 
\begin{eqnarray}
\sum_{k=0}^{l} (-1)^{l-k} \binom{L+1+l-k}{L}\binom{L+1}{k} = \binom{L+1}{l+1}
\label{eq4}
\end{eqnarray}
for $l=0,1,\cdots,n-1$. 
Consider the equality 
\begin{eqnarray}
(1+x)^{-(L+1)}(1+x)^{L+1}=1 
\label{eq5}
\end{eqnarray}
which holds for $|x|<1$. The left-hand side of (\ref{eq5}) can be written as 
\begin{eqnarray*}
\left( \sum_{p=0}^{\infty} (-1)^p \binom{L+p}{L} x^p \right) \left( \sum_{q=0}^{L+1} \binom{L+1}{q} x^q \right). 
\end{eqnarray*}
If we compare the coefficient of $x^{l+1}$ for $l=0,1,\cdots$ in both sides of the equality (\ref{eq5}), then we obtain 
\begin{eqnarray*}
\sum_{p+q=l+1} (-1)^p \binom{L+p}{L} \binom{L+1}{q}=0. 
\end{eqnarray*}
After a few algebras, we can derive the desired equality (\ref{eq4}). 

\hfill $\Box$ \\

Note that (\ref{eq3}) is identical to a closed-form expression 
for the {\it Eulerian number} $\left\langle {L \atop n-1} \right\rangle$ \cite{Graham1994}, where the Eulerian number 
$\left\langle {a \atop b} \right\rangle$ is the number of permutations $\pi$ of $\{1,\cdots,a\}$ that have exactly 
$b$ ascents, namely, $b$ places with $\pi(j)<\pi(j+1)$. The equality (\ref{eq2}) is equivalent to the so-called 
{\it Worpitzky's identity}: 
\begin{eqnarray}
n^L=\sum_{k=L-n}^{L-1} \left\langle {L \atop k} \right\rangle \binom{n+k}{L}. 
\label{eq6}
\end{eqnarray}
Indeed, one can obtain the Worpitzky's identity (\ref{eq6}) from (\ref{eq2}) by a few algebras using the symmetry law 
$\left\langle {L \atop i-1} \right\rangle = \left\langle {L \atop L-i} \right\rangle$. 

\subsection{The Minimal Realization Map $\mu$}
For any $\pi \in \mathcal{S}_L$, we can construct a word $s_1^L \in \mathbb{N}^L$ such that 
$\phi(s_1^L)=\pi$ in the following procedure: 
first, we decompose the sequence $\pi(1)\cdots\pi(L)$ into {\it maximal ascending sequences}. 
A subsequence $i_j \cdots i_{j+k}$ of a sequence $i_1 \cdots i_L$ is called a {\it maximal ascending 
sequence} if it is ascending, namely, $i_j \leq i_{j+1} \leq \cdots \leq i_{j+k}$, and neither 
$i_{j-1} i_j \cdots i_{j+k}$ nor $i_j \cdots i_{j+k} i_{j+k+1}$ is ascending. Suppose 
$\pi(1) \cdots \pi(i_1), \ \pi(i_1+1) \cdots \pi(i_2), \cdots, \pi(i_{k-1}+1) \cdots \pi(L)$ 
is a decomposition of $\pi(1)\cdots\pi(L)$ into maximal ascending sequences. 
If we define a word $s_1^L \in \mathbb{N}^L$ by 
\begin{eqnarray*}
s_{\pi(1)}=\cdots=s_{\pi(i_1)}=1, s_{\pi(i_1+1)}=\cdots=s_{\pi(i_2)}=2, \\ 
\cdots, s_{\pi(i_{k-1})+1}=\cdots=s_{\pi(L)}=k, 
\end{eqnarray*}
then we have $\phi(s_1^L)=\pi$ by construction. 
Thus, $\pi$ appears for the first time at most $k$. 
We denote the word $s_1^L$ by $\mu(\pi)$. $\mu$ defines a map 
$\mu : \mathcal{S}_L \to \mathbb{N}^L$ such that $\phi \circ \mu(\pi)=\pi$ for any $\pi \in \mathcal{S}_L$. 

For example, if $\pi \in \mathcal{S}_5$ is given by $\pi(1)\pi(2)\pi(3)\pi(4)\pi(5)=24315$, then its 
decomposition into maximal ascending sequences is $24,3,15$. If we put $s_2 s_4 s_3 s_1 s_5=11233$, then 
we obtain $\mu(\pi)=s_1 s_2 s_3 s_4 s_5=31213$. 

Let $\pi \in \mathcal{S}_L$ appear for the first time at $n$. By Lemma \ref{lem2}, 
there exists a unique word $s_1^L \in A_n^L$ such that $\phi(s_1^L)=\pi$. 
We say that $s_1^L$ is a {\it minimal realization} of $\pi$. 
In the following, we shall show that $\mu(\pi)$ is a minimal realization of $\pi$. 

\begin{Pro}
The following statements are equivalent:
\begin{itemize}
\item[(i)]
$s_1^L \in A_n^L$ is a minimal realization of some permutation $\pi \in \mathcal{S}_L$ that appears 
for the first time at $n$. 
\item[(ii)]
For any $1 \leq i \leq n-1$, there exists $1 \leq j < k \leq L$ such that $s_j=i+1, \ s_k=i$. 
\end{itemize}
\label{pro4}
\end{Pro}
{\it Proof.}
When $n=1$, the equivalence is trivial. So, we assume that $n \geq 2$ in the following discussion. 

(i)$\To$(ii): Let $s_1^L \in A_n^L$ be a minimal realization of $\pi \in \mathcal{S}_L$ that 
appears for the first time at $n$. 
Suppose that statement (ii) does not hold. Then, there exists $1 \leq i \leq n-1$ such that, 
for any $1 \leq j,k \leq n$, if $s_k=i$ and $s_j=i+1$, then $k<j$. 
Let us define a word $t_1^L$ by 
\begin{eqnarray*}
t_j=
\begin{cases}
s_j-1 & \text{if} \ s_j=i+1, \\
s_j & \text{otherwise}.
\end{cases}
\end{eqnarray*}
We claim that $\phi(t_1^L)=\pi$. By Corollary \ref{cor1}, it is sufficient to show that 
$s_k \leq s_j \Tot t_k \leq t_j$ for all $1 \leq k \leq j \leq L$. Fix $1 \leq k \leq j \leq L$. 
Assume that $s_k \leq s_j$. If $s_j=i+1$, then we have $t_j=s_j-1=i$. 
If we also have $s_k=i+1$, then $t_k=s_k-1=i=t_j$. Otherwise, we have $s_k \neq i+1$. 
Thus, we obtain $s_k \leq i$ because $s_k \leq s_j=i+1$. Then, $t_k=s_k \leq i=t_j$.
On the other hand, if $s_j \neq i+1$, then we have $t_j=s_j$. Thus, we obtain $t_k \leq s_k \leq s_j=t_j$. 
To show the reverse direction, let us assume that $t_k \leq t_j$. 
If $s_j=i+1$, then $t_j=s_j-1=i$. If we also have $s_k=i+1$, then $s_k=s_j$. Otherwise, we have $s_k \neq i+1$, 
then $t_k=s_k$ so that $s_k=t_k \leq t_j=i < s_j$. On the other hand, if $s_j \neq i+1$, then we have $t_j=s_j$. 
If we also have $s_k \neq i+1$, then $s_k=t_k \leq t_j=s_j$. Otherwise, we have $s_k=i+1$, then $t_k=s_k-1$. 
Suppose that $s_k > s_j$. Then, $s_j < s_k=i+1$ and $i=s_k-1=t_k \leq t_j=s_j$. Hence, $s_j=i$. Since 
we have assumed that (ii) does not hold, we obtain $j < k$. However, this contradicts our 
other assumption that $k \leq j$. Hence, we have $s_k \leq s_j$. 

Suppose that there exists $j$ such that $s_j=i+1$. Then, we have $t_1^L \neq s_1^L$. This contradicts 
the uniqueness of minimal realization of $\pi$ because both $s_1^L$ and $t_1^L$ are contained in 
$A_n^L$. Suppose that there exists no $j$ such that $s_j=i+1$. Since $\pi$ appears for the 
first time at $n$ and $s_1^L$ is its minimal realization, we have $s_{\pi(L)}=n$. Hence, 
$i+1<n$ should hold. Let us take the least $j$ such that $i+1 < s_{\pi(j)}$ and put it as $j_0$. 
If we define a word $t_1^L$ by 
\begin{eqnarray*}
t_{\pi(j)}=
\begin{cases}
s_{\pi(j)} & \text{if} \ j < j_0, \\
s_{\pi(j)}-1 & \text{if} \ j \geq j_0, 
\end{cases}
\end{eqnarray*}
then we have $\phi(t_1^L)=\pi$. 
Indeed, $t_{\pi(j_0-1)}=s_{\pi(j_0-1)}< i+1 \leq s_{\pi(j_0)}-1=t_{\pi(j_0)}$ 
because $i+1 < s_{\pi(j_0)}$. On the other hand, we have $t_1^L \in A_{n-1}^L$, which is a contradiction. 

(ii)$\To$(i): Assume that $s_1^L \in A_n^L$ satisfies (ii). Let $t_1^L \in A_{n-i}^L$ be 
a minimal realization of $\pi=\phi(s_1^L)$. We shall show that $i=0$. 
By Lemma \ref{lem2}, we have $t_{\pi(k)} \leq s_{\pi(k)}$ for $k=1,\cdots,L$ and 
$0 \leq c_{\pi(1)} \leq \cdots \leq c_{\pi(L)} = n-(n-i)=i$ for $c_k=s_k-t_k$. 
Suppose there exists $j$ such that $1 \leq c_{\pi(j)}$. Take the least $j$ such that 
$1 \leq c_{\pi(j)}$ and put it $j_0$. Now, consider the least $k$ such that 
$s_{\pi(k)}=s_{\pi(j_0)}$ and the largest $k'$ such that $s_{\pi(k')}=s_{\pi(j_0)}$ and 
put them as $k_0$ and $k_1$, respectively. Then, we have $t_{\pi(k_0)}=t_{\pi(k_1)}$. 
Indeed, $t_{\pi(k_0)}=s_{\pi(k_0)}-c_{\pi(k_0)}=s_{\pi(k_1)}-c_{\pi(k_0)} \geq s_{\pi(k_1)}-c_{\pi(k_1)}=t_{\pi(k_1)}$. 
On the other hand, $t_{\pi(k_0)} \leq t_{\pi(k_1)}$ because $k_0 \leq k_1$. 
Thus, we obtain $t_{\pi(k_0)} = t_{\pi(k_1)}$. This means that $c_{\pi(k_0)} = c_{\pi(k_1)}$, 
which, in turn, implies $c_{\pi(k)}=c_{\pi(j_0)}$ for all $k_0 \leq k \leq k_1$. 
(Thus, $j_0=k_0$. )
If we define a word $u_1^L$ by 
\begin{eqnarray*}
u_{\pi(k)}=
\begin{cases}
s_{\pi(k)}-1 & \text{if} \ k_0 \leq k \leq k_1, \\
s_{\pi(k)} & \text{otherwise}, 
\end{cases}
\end{eqnarray*}
then we have $\phi(u_1^L)=\pi$. To show this, we should care for only $k=k_0-1,k_0$ and $k=k_1,k_1+1$. 
First, let us consider the former. By the definition of 
$u_1^L$ and $k_0$, we have $u_{\pi(k_0-1)}=s_{\pi(k_0-1)}=t_{\pi(k_0-1)}$. 
We also have $u_{\pi(k_0)}=s_{\pi(k_0)}-1 \geq t_{\pi(k_0)}$ because 
$s_{\pi(k_0)}-t_{\pi(k_0)}=c_{\pi(k_0)} \geq 1$. Hence, 
$u_{\pi(k_0-1)}=t_{\pi(k_0-1)} \leq t_{\pi(k_0)} \leq u_{\pi(k_0)}$. 
If $u_{\pi(k_0-1)}=u_{\pi(k_0)}$, then $t_{\pi(k_0-1)}=t_{\pi(k_0)}$, which 
implies that $\pi(k_0-1) < \pi(k_0)$. 
The latter is obvious because $u_{\pi(k_1)}=s_{\pi(k_1)}-1 < s_{\pi(k_1+1)}=u_{\pi(k_1+1)}$. 

Now, if we put $s_{\pi(j_0)}=a(\geq 2)$, then there exist $j_1<j_2$ such that $s_{j_1}=a$ and $s_{j_2}=a-1$ 
by (ii). By the construction of $u_1^L$, we have $u_{j_1}=a-1=u_{j_2}$. This implies that 
$u_1^L$ and $s_1^L$ have different rank sequences because $\varphi(u_1^L)_{j_2}>\varphi(s_1^L)_{j_2}$. 
Thus, we have $\phi(u_1^L)=\iota \circ \varphi(u_1^L) \neq \iota \circ \varphi(s_1^L)=\phi(s_1^L)$, 
which is a contradiction. 

\hfill $\Box$ \\

\begin{Cor}
For $\pi \in \mathcal{S}_L$, $\mu(\pi)$ is a minimal realization of $\pi$. 
\label{cor2}
\end{Cor}
{\it Proof.}
Let 
\begin{eqnarray*}
\pi(1) \cdots \pi(j_1), \ \pi(j_1+1) \cdots \pi(j_2), \cdots, \pi(j_{k-1}+1) \cdots \pi(L) 
\end{eqnarray*}
be a decomposition of $\pi(1)\cdots\pi(L)$ into maximal ascending sequences. 
If $s_1^L=\mu(\pi)$, then 
\begin{eqnarray*}
s_{\pi(1)}=\cdots=s_{\pi(j_1)}=1, s_{\pi(j_1+1)}=\cdots=s_{\pi(j_2)}=2, \\ 
\cdots, s_{\pi(j_{k-1})+1}=\cdots=s_{\pi(L)}=k
\end{eqnarray*}
by the definition of $\mu$. For each $1 \leq i \leq k-1$, we have 
$s_{\pi(j_i)}=i$, $s_{\pi(j_i+1)}=i+1$ and $\pi(j_i) > \pi(j_i+1)$. Hence, condition (ii) 
of Proposition \ref{pro4} is satisfied by $s_1^L$. Since $\phi(s_1^L)=\pi$, $s_1^L$ is a minimal 
realization of $\pi$. 

\hfill $\Box$ \\

\subsection{The Duality}
We can make the pair of maps 
\begin{eqnarray*}
\xymatrix{
\mathbb{N}^L \ar@<1ex>[r]^{\phi} & \mathcal{S}_L \ar@<1ex>[l]^{\mu}
}
\end{eqnarray*}
form a Galois connection \cite{Davey2002} in the following way: 
we consider the set $\mathcal{S}_L$ as an ordered set with the discrete order, namely, 
we define an order relation $\leq_{\mathcal{S}_L}$ on $\mathcal{S}_L$ by 
$\pi \leq_{\mathcal{S}_L} \pi' :\Tot \pi=\pi'$. On the other hand, we introduce 
an order relation $\leq_{\mathbb{N}^L}$ on $\mathbb{N}^L$ by 
$s_1^L \leq_{\mathbb{N}^L} t_1^L :\Tot \phi(s_1^L)=\phi(t_1^L)=:\pi$ and 
there exist $0 \leq c_{\pi(1)} \leq \cdots \leq c_{\pi(L)}$ such that $s_k=c_k+t_k$. 
By Corollary \ref{cor2}, we have 
\begin{eqnarray*}
\phi(s_1^L) \leq_{\mathcal{S}_L} \pi \Tot s_1^L \leq_{\mathbb{N}^L} \mu(\pi)
\end{eqnarray*}
for $s_1^L \in \mathbb{N}^L$ and $\pi \in \mathcal{S}_L$. 

If we restrict the domain of the map $\phi$ to $A_n^L$, we obtain the 
following form of the duality stated in Theorem \ref{thm1} (iv) bellow. 
Theorem \ref{thm1} summarizes the main results of this section. 

\begin{Thm}
Let us set the domain of the coarse-graining map $\phi$ to $A_n^L$. 
\begin{itemize}
\item[(i)]
For $\pi \in {\mathcal S}_L$, 
if $\phi^{-1}(\pi) \neq \emptyset$, then the value of $|\phi^{-1}(\pi)|$ takes 
a binomial coefficient $\binom{L+n-i}{n-i}$ for some $1 \leq i \leq n$. 
\item[(ii)]
For $\pi \in {\mathcal S}_L$, the following two statements are equivalent: 
\begin{itemize}
\item[(a)]
$|\phi^{-1}(\pi)|=1$. 
\item[(b)]
$\pi$ appears for the first time at $n$. 
\end{itemize}
\item[(iii)]
For $s_1^L \in A_n^L$, the following three statements are equivalent: 
\begin{itemize}
\item[(c)]
$\phi^{-1}(\pi)=\{s_1^L\}$ for some $\pi \in {\mathcal S}_L$. 
\item[(d)]
For any $1 \leq i \leq n-1$ there exists $1 \leq j < k \leq L$ such that 
$s_j=i+1,s_k=i$. 
\item[(e)]
$s_1^L \not \in A_{n-1}^L$ and $s_1^L=\mu \circ \phi(s_1^L)$. 
\end{itemize}
\item[(iv)]
If we restrict $\phi$ on the subset of $A_n^L$ consisting of words satisfying one of the three 
equivalent conditions in (iii), then $\phi$ gives a one-to-one correspondence between these words 
and permutations of length $L$ satisfying one of the two equivalent conditions in (ii) with its 
inverse $\mu$. 
\end{itemize}
\label{thm1}
\end{Thm}
{\it Proof.}
(i) If $\pi$ appears for the first time at $i \leq n$, then $|\phi^{-1}(\pi)|= \binom{L+n-i}{n-i}$ 
by Lemma \ref{lem2} (ii). 

(ii) (a)$\To$(b): Suppose $|\phi^{-1}(\pi)|=1$ and $\pi$ appears for the first time at $i \leq n$. 
By (i), $\binom{L+n-i}{n-i}=1$ holds. This happens if and only if $i=n$. 

(b)$\To$(a): If $\pi$ appears for the first time at $n$, then there exists a unique $s_1^L \in A_n^L$ 
such that $\phi(s_1^L)=\pi$. Hence, $\phi^{-1}(\pi)=\{s_1^L\}$. 

(iii) (c)$\To$(d),(e): Assume $\phi^{-1}(\pi)=\{s_1^L\}$ for some $\pi \in {\mathcal S}_L$. 
By (ii), $\pi$ appears for the first time at $n$. Hence, $s_1^L$ is a minimal realization of $\pi$. 
Hence, (d) holds for $s_1^L$ by Proposition \ref{pro4}. To see (e) holds, first observe that 
$s_1^L$ cannot be contained in $A_{n-1}^L$. We also have $s_1^L=\mu(\pi)=\mu(\phi(s_1^L))$ because 
$\mu(\pi)$ is a minimal realization of $\pi$ by Corollary \ref{cor2}. 

(d)$\To$(c): If (d) holds for $s_1^L$, then $s_1^L$ is a minimal realization of some $\pi \in {\mathcal S}_L$ 
that appears for the first time at $n$ by Proposition \ref{pro4}. 
Hence, we have $\phi^{-1}(\pi)=\{s_1^L\}$ by the uniqueness of minimal realization. 

(e)$\To$(c): Assume $s_1^L \not\in A_{n-1}^L$ and $s_1^L=\mu(\phi(s_1^L))$. 
$s_1^L$ is a minimal realization of $\phi(s_1^L)$ by Corollary \ref{cor2}. 
$\phi(s_1^L)$ appears for the first time at $n$ since $s_1^L \not\in A_{n-1}^L$. 
Hence, $\phi^{-1}(\phi(s_1^L))=\{s_1^L\}$ holds by (ii). 

(iv) Let us put $X=\{s_1^L \in A_n^L | s_1^L \not\in A_{n-1}^L, \ s_1^L=\mu \circ \phi(s_1^L) \}$ and 
$Y=\{\pi \in {\mathcal S}_L | |\phi^{-1}(\pi)|=1 \}$. If $s_1^L \in X$, then 
$\phi^{-1}(\phi(s_1^L))=\{s_1^L\}$. Hence, $\phi$ restricted on $X$ is a map from $X$ into $Y$. 
On the other hand, $\mu$ restricted on $Y$ is a map from $Y$ into $X$. Indeed, 
$\pi$ appears for the first time at $n$ by (ii). Since $\mu(\pi)$ is a minimal realization 
of $\pi$ by Corollary \ref{cor2}, it must hold that $\phi^{-1}(\pi)=\{\mu(\pi)\}$. 
Thus, we have $\mu(\pi) \not\in A_{n-1}^L$ and $\mu(\pi) \in A_n^L$. 
We also have $\mu(\pi)=\mu \circ \phi \circ \mu(\pi)$ because $\phi \circ \mu$ is an identity on ${\mathcal S}_L$. 
Now, $\mu$ restricted on $Y$ is a left inverse of $\phi$ restricted on $X$ by the definition of $X$. 
It is also a right inverse because $\phi \circ \mu$ is an identity on ${\mathcal S}_L$. 
\hfill $\Box$ \\

\section{Permutation Entropy Rate Revisited}
Let ${\bf S}=\{S_1,S_2,\cdots \}$ be a finite-state stationary stochastic process, where 
stochastic variables $S_i$ take their values in $A_n$. Stationarity means that 
\begin{eqnarray*}
{\rm Pr}\{S_1=s_1,\cdots,S_L=s_L\}={\rm Pr}\{S_{k+1}=s_1,\cdots,S_{k+L}=s_L\}
\end{eqnarray*}
for any $k, L \geq 1$ and $s_1,\cdots,s_L \in A_n$. For simplicity, we write 
$p(s_1^L)=p(s_1 \cdots s_L)$ instead of ${\rm Pr}\{S_1=s_1,\cdots,S_L=s_L\}$. 
In the following discussion, we set the domain of the map $\phi$ introduced in Section 2 to $A_n^L$. 

The {\it entropy rate} $h({\bf S})$ of a finite-state stationary stochastic process 
${\bf S}=\{S_1,S_2,\cdots \}$ is defined by 
\begin{eqnarray}
h({\bf S})=\lim_{L \to \infty} \frac{1}{L} H(S_1^L), 
\label{eq7}
\end{eqnarray}
where $H(S_1^L)=H(S_1,\cdots,S_L)=-\sum_{s_1^L \in A_n^L} p(s_1^L) \log p(s_1^L)$. 
Here, we take the base of the logarithm as 2. It is well-known that the limit exists for any 
finite-state stationary stochastic process \cite{Cover1991}. 

The {\it permutation entropy rate} $h^*({\bf S})$ of a finite-state stationary stochastic process 
${\bf S}=\{S_1,S_2,\cdots \}$ is defined by 
\begin{eqnarray}
h^*({\bf S})=\lim_{L \to \infty} \frac{1}{L} H^*(S_1^L), 
\label{eq8}
\end{eqnarray}
where $H^*(S_1^L)=H^*(S_1,\cdots,S_L)=-\sum_{\pi \in \mathcal{S}_L} p(\pi) \log p(\pi)$ and 
$p(\pi)$ is the probability that $\pi$ is realized in ${\bf S}$, namely, $p(\pi)=\sum_{s_1^L \in \phi^{-1}(\pi)} p(s_1^L)$ for $\pi \in \mathcal{S}_L$. 
Amig\'o et al. proved that the limit exists for all finite-state stationary stochastic processes and is equal to $h({\bf S})$ \cite{Amigo2010,Amigo2005}. 
They first showed the equality with the assumption of the ergodicity. Then, they proceeded to the general case by appealing to 
the ergodic decomposition theorem of the entropy rate. 

If we make use of {\it rank variables} $R_i= \sum_{j=1}^n \delta \left( S_j \leq S_i \right)$ for $i=1,2,\cdots$ 
introduced in $\cite{Amigo2005}$, then the permutation entropy rate has the following alternative expression by 
Proposition \ref{pro1} and Proposition \ref{pro2}: 
\begin{eqnarray*}
h^*({\bf S})=\lim_{L \to \infty} \frac{1}{L} H(R_1^L). 
\end{eqnarray*}

Intuitively, the entropy rate quantifies the uncertainty of values per unit symbol on the one hand, 
while the permutation entropy rate quantifies the uncertainty of orderings between values per unit symbol on the other hand. 

In the following discussion, we give an elementary alternative proof of $h({\bf S})=h^*({\bf S})$ for a finite-state stationary stochastic 
process ${\bf S}=\{S_1,S_2,\cdots \}$ based on the duality between values and orderings established in Section 2. 

\begin{Lem}
\begin{eqnarray}
0 \leq H(S_1^L)-H^*(S_1^L) \leq 
\left( \sum_{\begin{subarray}{c} \pi \in \mathcal{S}_L, \\ |\phi^{-1}(\pi)|>1 \end{subarray} } p(\pi) \right) n \log (L+n). 
\label{eq9}
\end{eqnarray}
\label{lem3}
\end{Lem}
{\it Proof.}
\begin{eqnarray*}
H(S_1^L)-H^*(S_1^L) 
&=& -\sum_{s_1^L \in A_n^L} p(s_1^L) \log p(s_1^L)+\sum_{\pi \in \mathcal{S}_L} p(\pi) \log p(\pi) \\
&=& \sum_{\pi \in \mathcal{S}_L} 
\left( -\sum_{ s_1^L \in \phi^{-1}(\pi) } p(s_1^L) \log p(s_1^L) 
+\left( \sum_{ s_1^L \in \phi^{-1}(\pi) } p(s_1^L) \right) \log p(\pi)
\right) \\
&=& \sum_{\begin{subarray}{c} \pi \in \mathcal{S}_L, \\ p(\pi)>0 \end{subarray} }
\left( -\sum_{ s_1^L \in \phi^{-1}(\pi) } p(s_1^L) \log \frac{p(s_1^L)}{p(\pi)} 
\right) \\
&=& \sum_{\begin{subarray}{c} \pi \in \mathcal{S}_L, \\ p(\pi)>0 \end{subarray} } p(\pi) 
\left( -\sum_{ s_1^L \in \phi^{-1}(\pi) } \frac{p(s_1^L)}{p(\pi)} \log \frac{p(s_1^L)}{p(\pi)} 
\right).
\end{eqnarray*}
Now, we have 
\begin{eqnarray*}
0 \leq -\sum_{ s_1^L \in \phi^{-1}(\pi) } \frac{p(s_1^L)}{p(\pi)} \log \frac{p(s_1^L)}{p(\pi)} 
\leq n \log (L+n)
\end{eqnarray*}
for $\pi \in \mathcal{S}_L$ such that $\phi^{-1}(\pi) \neq \emptyset$ and $p(\pi)>0$ because the value of $|\phi^{-1}(\pi)|$
takes a binomial coefficient $\binom{L+n-i}{n-i}$ for some $1 \leq i \leq n$ by Theorem \ref{thm1} (i). Note that if 
$i=n$, then $|\phi^{-1}(\pi)|=1$, which implies 
\begin{eqnarray*}
-\sum_{ s_1^L \in \phi^{-1}(\pi) } \frac{p(s_1^L)}{p(\pi)} \log \frac{p(s_1^L)}{p(\pi)} 
=0.
\end{eqnarray*}

\hfill $\Box$ \\

\begin{Thm}
For any finite-state stationary stochastic process ${\bf S}=\{S_1,S_2,\cdots \}$, $h({\bf S})=h^*({\bf S})$. 
\label{thm3}
\end{Thm}
{\it Proof.}
Since we have 
\begin{eqnarray*}
\sum_{\begin{subarray}{c} \pi \in \mathcal{S}_L, \\ |\phi^{-1}(\pi)|>1 \end{subarray} } p(\pi) \leq 1 \ \text{and} 
\ \frac{\log(L+n)}{L} \underset{L \to \infty}{\to} 0, 
\end{eqnarray*}
we obtain 
\begin{eqnarray*}
h^*({\bf S})=\lim_{L \to \infty} \frac{H^*(S_1^L)}{L}=\lim_{L \to \infty} \frac{H(S_1^L)}{L}=h({\bf S})
\end{eqnarray*}
by Lemma \ref{lem3}. 

\hfill $\Box$ \\

\section{Permutation Excess Entropy}
The {\it excess entropy} \cite{Crutchfield2003} ${\bf E}({\bf S})$ of a finite-state stationary stochastic process 
${\bf S}=\{S_1,S_2,\cdots \}$ is defined by 
\begin{eqnarray}
{\bf E}({\bf S})=\lim_{L \to \infty} \left( H(S_1^L) - h({\bf S})L \right), 
\label{eq10}
\end{eqnarray}
if the limit on the right-hand side exists. The excess entropy ${\bf E}({\bf S})$ is a measure of 
global correlation present in a finite-state stationary stochastic process ${\bf S}=\{S_1,S_2,\cdots \}$. 
If ${\bf E}({\bf S})$ exists, then we can write \cite{Crutchfield2003}
\begin{eqnarray}
{\bf E}({\bf S})=\sum_{L=1}^{\infty} \left( H(S_L|S_1^{L-1}) - h({\bf S}) \right)=\lim_{L \to \infty} I(S_1^L ; S_{L+1}^{2L}), 
\label{eq11}
\end{eqnarray}
where $H(Y|X)$ is the conditional entropy of $Y$ given $X$ and $I(X;Y)$ is the mutual information between $X$ and $Y$ 
for stochastic variables $X$ and $Y$. 

The {\it permutation excess entropy} ${\bf E}^*({\bf S})$ of a finite-state stationary stochastic process 
${\bf S}=\{S_1,S_2,\cdots \}$ is defined by 
\begin{eqnarray}
{\bf E}^*({\bf S})=\lim_{L \to \infty} \left( H^*(S_1^L) - h^*({\bf S})L \right), 
\label{eq12}
\end{eqnarray}
if the limit on the right-hand side exists. 

It is straightforward to obtain a similar alternative expression for the permutation excess entropy ${\bf E}^*({\bf S})$ 
to that for the excess entropy (\ref{eq11}), when ${\bf E}^*({\bf S})$ exists: 
\begin{eqnarray}
{\bf E}^*({\bf S})=\sum_{L=1}^{\infty} \left( H(R_L|R_1^{L-1}) - h^*({\bf S}) \right). 
\label{eq13}
\end{eqnarray}
Note that we also have the equality $h^*({\bf S})=\lim_{L \to \infty} H(R_L|R_1^{L-1})$ 
which is an analog to the alternative expression for the entropy rate 
$h({\bf S})=\lim_{L \to \infty} H(S_L|S_1^{L-1})$ because the right-hand side expression in (\ref{eq13}) converges. 
We can prove that the permutation excess entropy ${\bf E}^*({\bf S})$ also admits a mutual information expression if 
the process ${\bf S}$ is ergodic Markov, which will be presented elsewhere \cite{Haruna2011}. 

We would like to know whether ${\bf E}({\bf S})={\bf E}^*({\bf S})$ holds or not for a given finite-state 
stationary stochastic process ${\bf S}$. In the rest of the paper, we give a partial answer to this problem. 
In particular, we will show that ${\bf E}({\bf S})={\bf E}^*({\bf S})$ for any finite-state stationary ergodic Markov process. 

Note that we always have ${\bf E}^*({\bf S}) \leq {\bf E}({\bf S})$ if the limits on both sides exist 
because $H^*(S_1^L) \leq H(S_1^L)$  and $h^*({\bf S})=h({\bf S})$ by Lemma \ref{lem3} and Theorem \ref{thm3}, 
respectively. To show ${\bf E}^*({\bf S}) = {\bf E}({\bf S})$, it is sufficient to show that 
\begin{eqnarray*}
\left( \sum_{\begin{subarray}{c} \pi \in \mathcal{S}_L, \\ |\phi^{-1}(\pi)|>1 \end{subarray} } p(\pi) \right) \log L 
\underset{L \to \infty}{\to} 0 
\end{eqnarray*}
if ${\bf E}({\bf S})$ exists. 
Let us put 
\begin{eqnarray*}
q_L:=\sum_{\begin{subarray}{c} \pi \in \mathcal{S}_L, \\ |\phi^{-1}(\pi)|>1 \end{subarray} } p(\pi). 
\end{eqnarray*}

\begin{Lem}
Let $\epsilon$ be a positive real number and $L$ be a positive integer. 
Assume that for any $s \in A_n$, 
\begin{eqnarray*}
{\rm Pr}\{ s_1^{\lfloor L/2 \rfloor} | s_j \neq s \text{ for any } 1 \leq j \leq \lfloor L/2 \rfloor \}
\leq \epsilon 
\end{eqnarray*}
holds, where $\lfloor x \rfloor$ is the largest integer not greater than $x$. Then, we have $q_L \leq 2n \epsilon$. 
\label{lem4}
\end{Lem}
{\it Proof.}
We shall prove 
\begin{eqnarray*}
\sum_{\begin{subarray}{c} \pi \in \mathcal{S}_L, \\ |\phi^{-1}(\pi)|=1 \end{subarray} } p(\pi) \geq 1-2n \epsilon. 
\end{eqnarray*}
Let us consider a word $s_1^L \in A_n^L$ satisfying the following two conditions: 
\begin{itemize}
\item[(i)]
Each symbol $s \in A_n$ appears in $s_1^{\lfloor L/2 \rfloor}$ at least once. 
\item[(ii)]
Each symbol $s \in A_n$ appears in $s_{\lfloor L/2 \rfloor+1}^L$ at least once. 
\end{itemize}
By the assumption of the lemma, we have 
\begin{eqnarray*}
{\rm Pr}\{ s_1^{\lfloor L/2 \rfloor} | \text{(i) holds} \} \geq 1-n\epsilon, 
\end{eqnarray*}
because 
\begin{eqnarray*}
{\rm Pr}\{ s_1^{\lfloor L/2 \rfloor} | \text{(i) holds} \} 
+ \sum_{s=1}^n {\rm Pr}\{ s_1^{\lfloor L/2 \rfloor} | s_j \neq s \text{ for any } 1 \leq j \leq \lfloor L/2 \rfloor \} \geq 1. 
\end{eqnarray*}
Similarly, 
\begin{eqnarray*}
{\rm Pr}\{ s_{\lfloor L/2 \rfloor + 1}^L | \text{(ii) holds} \} \geq 1-n\epsilon 
\end{eqnarray*}
holds because of the stationarity. Hence, we have both 
\begin{eqnarray*}
{\rm Pr}\{ s_1^L | \text{(i) holds} \} \geq 1-n\epsilon \text{ and }
{\rm Pr}\{ s_1^L | \text{(ii) holds} \} \geq 1-n\epsilon, 
\end{eqnarray*}
which imply 
\begin{eqnarray*}
{\rm Pr}\{ s_1^L | \text{both (i) and (ii) hold} \} \geq 1-2n\epsilon. 
\end{eqnarray*}
It is clear that a word $s_1^L \in A_n^L$ satisfying both (i) and (ii) fulfills 
condition (d) in Theorem \ref{thm1} (iii). 
Hence, by Theorem \ref{thm1} (iv), we obtain
\begin{eqnarray*}
\sum_{\begin{subarray}{c} \pi \in \mathcal{S}_L, \\ |\phi^{-1}(\pi)|=1 \end{subarray} } p(\pi) 
=\textstyle{\sum^*} p(s_1^L)
\geq {\rm Pr}\{ s_1^L | \text{both (i) and (ii) hold} \} \geq 1-2n\epsilon, 
\end{eqnarray*}
where $\sum^*$ is the sum over all words $s_1^L$ satisfying the condition (d) in Theorem \ref{thm1} (iii). 

\hfill $\Box$ \\

As a first simple application of Lemma \ref{lem4}, let us consider a stochastic process 
${\bf S}=\{S_1,S_2,\cdots\}$ such that the stochastic variables $S_i$ are independent and 
identically distributed, namely, each symbol $s \in A_n$ appears at a probability $p(s)>0$ 
independently. If we put $0<\alpha:=\min_{s \in A_n} \{ p(s) \}<1$, then we have 
\begin{eqnarray*}
{\rm Pr}\{ s_1^{\lfloor L/2 \rfloor} | s_j \neq s \text{ for any } 1 \leq j \leq \lfloor L/2 \rfloor \}
=\left( 1-p(s) \right)^{\lfloor L/2 \rfloor} \leq \left\{ ( 1-\alpha)^\frac{1}{2} \right\}^L. 
\end{eqnarray*}
Thus, by Lemma \ref{lem4}, we have 
\begin{eqnarray*}
H(S_1^L)-H^*(S_1^L) \leq 2n^2 \left\{ ( 1-\alpha)^\frac{1}{2} \right\}^L \log(L+n) \underset{L \to \infty}{\to} 0. 
\end{eqnarray*}
However, in this case, ${\bf E}^*({\bf S})={\bf E}({\bf S})$ is obvious from 
${\bf E}^*({\bf S}) \leq {\bf E}({\bf S})$ because ${\bf E}({\bf S})=0$. 

Let ${\bf S}=\{S_1,S_2,\cdots\}$ be a {\it finite-state stationary ergodic Markov process} with a set of states $A_n$ and 
a transition matrix $P=(p_{ij})$, where $p_{ij} \geq 0$ for all $1 \leq i,j \leq n$ and $\sum_{j=1}^n p_{ij}=1$ for 
all $1 \leq i \leq n$. 
It is known that a finite-state stationary Markov process is ergodic if and only if its transition matrix 
$P$ is irreducible \cite{Walters1982}: a matrix $P$ is {\it irreducible} if for all $1 \leq i,j \leq n$ there exists $l>0$ 
such that $p_{ij}^{(l)}>0$, where $p_{ij}^{(l)}$ is the $(i,j)$-th element of $P^l$. By the Perron-Frobenius theorem 
for irreducible non-negative matrices, there exists a unique {\it stationary distribution} 
${\bf p}=(p_1,\cdots,p_n)$ such that $p_i>0$ for all $1 \leq i \leq n$, $\sum_{i=1}^n p_i=1$ 
and $\sum_{i=1}^n p_i p_{ij}=p_j$ for all $1 \leq j \leq n$, namely, 
$^tP{\bf p}={\bf p}$, where $^tP$ is the transpose of the matrix $P$. Then, we have 
$p(s_1^L)=p_{s_1} p_{s_1 s_2} \cdots p_{s_{L-1} s_L}$ for $s_1^L \in A^L$. The entropy rate $h({\bf S})$ and 
the excess entropy ${\bf E}({\bf S})$ of a finite-state stationary Markov process ${\bf S}=\{S_1,S_2,\cdots\}$ are given by 
$h({\bf S})=-\sum_{i,j=1}^n p_i p_{ij} \log p_{ij}$ and 
${\bf E}({\bf S})=-\sum_{i=1}^n p_i \log p_i + \sum_{i,j=1}^n p_i p_{ij} \log p_{ij}$, respectively. 

Let $L$ be a positive integer. Let us put $N:=\lfloor L/2 \rfloor$. Given a symbol $s \in A_n$, 
we would like to evaluate 
\begin{eqnarray*}
\beta_s &:=& {\rm Pr}\{ s_1^N | s_j \neq s \text{ for any } 1 \leq j \leq N \} \\
&=& \sum_{\begin{subarray}{c} s_j \neq s, \\ 1 \leq j \leq N \end{subarray} } p(s_1 \cdots p_N) 
= \sum_{\begin{subarray}{c} s_j \neq s, \\ 1 \leq j \leq N \end{subarray}} p_{s_1} p_{s_1 s_2} \cdots p_{s_{N-1} s_N}. 
\end{eqnarray*}
If $n=1$ then $\beta_1=0$. So, this case is trivial. Hence, we assume $n \geq 2$ in the following discussion. 
If we introduce a matrix $P_s$ whose $(i,j)$-th elements are defined by 
\begin{eqnarray*}
\left( P_s \right)_{ij}=
\begin{cases}
0 & \text{ if } i=s \\
p_{ij} & \text{otherwise,}
\end{cases}
\end{eqnarray*}
then we can write 
\begin{eqnarray*}
\beta_s=\langle \left(P_s \right)^{N-1} {\bf u}_s, {\bf p} \rangle, 
\end{eqnarray*}
where a vector ${\bf u}_s=(u_1,\cdots,u_n)$ is defined by $u_i=0$ if $i=s$ otherwise $u_i=1$ and 
$\langle \cdots,\cdots \rangle$ is the usual inner product in the $n$-dimensional Euclidean space. 

Since $P_s$ is a non-negative matrix, the following statements hold by the Perron-Frobenius theorem 
for non-negative matrices: 
\begin{itemize}
\item[(i)]
There exists a non-negative eigenvalue $\lambda$ such that 
any other eigenvalue of $P_s$ has absolute value not greater than $\lambda$. 
\item[(ii)]
$\lambda \leq \max_{i} \{ \sum_{j=1}^n \left( P_s \right)_{ij} \}=1$. 
\item[(iii)]
There exists a non-negative right eigenvector ${\bf v}$ corresponding to the eigenvalue $\lambda$. 
\end{itemize}

\begin{Lem}
$\lambda < 1$. 
\label{lem5}
\end{Lem}
{\it Proof.}
Suppose that $\lambda=1$. Then, we have $P_s {\bf v}={\bf v}$. 
For any positive integer $l$, we have 
\begin{eqnarray*}
\langle {\bf v}, {\bf p} \rangle 
=\langle P_s^l{\bf v}, {\bf p} \rangle 
\leq \langle P^l{\bf v}, {\bf p} \rangle 
=\langle {\bf v}, \left( ^tP \right)^l{\bf p} \rangle 
=\langle {\bf v}, {\bf p} \rangle, 
\end{eqnarray*}
since $P_s \leq P$. Thus, we obtain $\langle \left(P^l-P_s^l \right){\bf v}, {\bf p} \rangle=0$, which 
implies that $\left(P^l-P_s^l \right){\bf v}={\bf 0}$ because ${\bf p}$ is a positive vector and 
$\left(P^l-P_s^l \right){\bf v}$ is a non-negative vector. 
Now, let us fix any $1 \leq j \leq n$. There exists $l$ such that $p_{sj}^{(l)}>0$ because $P$ is irreducible. 
Since the elements in the $s$-th row of the matrix $P_s^l$ are all $0$, 
we have $\sum_{k=1}^n p_{sk}^{(l)} v_k=0$, where we put ${\bf v}=(v_1,\cdots,v_n)$. Thus, we obtain 
$v_j=0$ because $p_{sj}^{(l)}>0$, $p_{sk}^{(l)} \geq 0$ and $v_k \geq 0$ for all $1 \leq k \leq n$.
Since $1 \leq j \leq n$ is arbitrary, ${\bf v}={\bf 0}$ must hold. However, this contradicts ${\bf v} \neq {\bf 0}$ because ${\bf v}$ is an eigenvector. 

\hfill $\Box$ \\

Now, let $P_s=S+T$ be a Jordan-Chevalley decomposition of the matrix $P_s$, where $S$ is a diagonalizable matrix 
and $T$ is a nilpotent matrix. Let $A$ be an invertible matrix such that $A^{-1}SA=D$, where $D$ is 
a diagonal matrix. Since $T$ is nilpotent, there exists a positive integer $k$ such that $T^k$ is a zero matrix. 
We also have $ST=TS$. If we put $E:=A^{-1}TA$ then $E^k$ is a zero matrix and $DE=ED$. 
Thus, for sufficiently large $N$, 
\begin{eqnarray*}
P_s^{N-1} &=& A(D+E)^N A^{-1} \\
&=& A \left( \sum_{i=0}^{k-1} \binom{N-1}{i} D^{N-1-i} E^i \right) A^{-1}=\lambda^{N-k} O(N^{k-1}), 
\end{eqnarray*}
where the big-$O$ notation $O(N^{k-1})$ for a matrix means that each element of the matrix is $O(N^{k-1})$. 
Hence, we obtain $\beta_s=\lambda^{N-k} O(N^{k-1})$. Since $0 \leq \lambda < 1$ by Lemma \ref{lem5}, 
we get the following theorem by combining Lemma \ref{lem3} and Lemma \ref{lem4}: 

\begin{Thm}
Let ${\bf S}=\{S_1,S_2,\cdots\}$ is a finite-state stationary ergodic Markov process. 
Then, the permutation excess entropy ${\bf E}^*({\bf S})$ exists and 
${\bf E}^*({\bf S})={\bf E}({\bf S})$. 
\label{thm2}
\end{Thm}

We can construct a finite-state stationary non-ergodic Markov process such that ${\bf E}({\bf S}) \neq {\bf E}^*({\bf S})$ 
immediately. For example, let $n=2$ and 
\begin{eqnarray*}
P=
\begin{pmatrix}
1 & 0 \\
0 & 1
\end{pmatrix}.
\end{eqnarray*}
We choose a stationary distribution 
${\bf p}=(p_1,p_2)=(\frac{1}{2},\frac{1}{2})$. Then we have 
$p(\underbrace{00 \cdots 0}_{L})=p(\underbrace{11 \cdots 1}_{L})=\frac{1}{2}$. Hence we have 
$h({\bf S})=h^*({\bf S})=0$ and ${\bf E}({\bf S})=-p_1 \log p_1 - p_2 \log p_2 =1$. 
On the other hand, we have ${\bf E}^*({\bf S})=0$ because 
$\phi(\underbrace{00 \cdots 0}_{L})=\phi(\underbrace{11 \cdots 1}_{L}) \in \mathcal{S}_L$. 

\subsection*{Acknowledgments}
TH was supported by the JST PRESTO program.

\end{document}